\documentclass[amssymb,prb,twocolumn,showpacs,superscriptaddress,reprint]{revtex4} 
\usepackage{amsmath}
\usepackage{bbold}
\usepackage{amsbsy}
\usepackage{graphicx}
\usepackage{verbatim}
\usepackage{graphics}

\newcommand{\bea}{\begin{eqnarray}}
\newcommand{\eea}{\end{eqnarray}}
\newcommand{\be}{\begin{equation}}
\newcommand{\ee}{\end{equation}}

\begin{document}

\title{Detection of single electron heat transfer statistics}
\author{Rafael S\'anchez}
\affiliation{Instituto de Ciencia de Materiales de Madrid (ICMM-CSIC), Cantoblanco 28049 Madrid, Spain}
\author{Markus B\"uttiker}
\affiliation{D\'epartement de Physique Th\'eorique, Universit\'e de Gen\`eve, CH-1211 Gen\`eve 4, Switzerland}

\begin{abstract}
We consider a quantum dot system whose charge fluctuations are monitored by a quantum point contact allowing for the detection of both charge and transferred heat statistics. Our system consists of two nearby conductors that exchange energy via Coulomb interaction. In interfaces consisting of capacitively coupled quantum dots, energy transfer is discrete. In the presence of hot spots, we predict gate dependent deviations away from a fluctuation theorem that holds for charge currents in the absence of temperature gradients. A fluctuation theorem holds for coupled dot configurations with heat exchange and no net particle flow.
\end{abstract}
\maketitle

\section{Introduction}
Contrary to charge, heat currents are difficult to measure on nanoscopic scales~\cite{giazotto}. Seebeck-Onsager coefficients relate heat flows to voltage drops~\cite{butcher} allowing for the detection of mesoscopic heat currents in the linear regime~\cite{molenkamp, molenkamp2}. Recently multiterminal arrangements have been used to inject current and as thermometers in heat transistors~\cite{saira}. However, a method for a direct measurement of non linear electronic heat currents and their fluctuations has not been described. It would allow for instance to investigate recent proposals of refrigerators~\cite{edwards,smith}, efficient  heat converters~\cite{humphrey,hotspots,muralidharan,bjorn2}, rectifiers~\cite{scheibner} or diodes~\cite{ruokola} based on single electron transistors. 

Measurements of the charge counting statistics have become a standard way to access the noise properties of the electric flows through quantum dot systems~\cite{gustavsson,fujisawa,eugene}. There, time resolved charge fluctuations are detected by current changes though a capacitively coupled quantum point contact (QPC). Experimental precision makes possible to measure the behaviour of high order cumulants or finite frequency statistics~\cite{christian,ubbelohde}. For this reason, mesoscopic conductors offer an ideal playground for experimental verifications of nonlinear fluctuation relations as expressed for electric conduction~\cite{tobiska,andrieux,saito-heat,heidi,saito,rosa}. 
Pioneering experiments have been carried out recently to test work fluctuation relations (Jarzynsky and Crooks) in driven closed conductors~\cite{averinpekola,saira2} as well as in open conductors~\cite{nakamura,kung} of interest here.  In experiments where the detector is asymmetrically coupled to a double quantum dot~\cite{fujisawa,kung}, different charge distributions are resolved as different outputs in the QPC. This way, trajectories of charges flowing along and against an applied voltage can be traced out. If the external force is thermal, so that temperature is inhomogeneous, the fluctuation theorem for charge currents does not hold. One must then consider both charge and energy currents~\cite{andrieux2} or, alternatively, heat. To test fluctuation relations, the ability to measure heat currents is therefore also of fundamental importance. 

\begin{figure}[b]
\center
\includegraphics[width=88mm,clip]{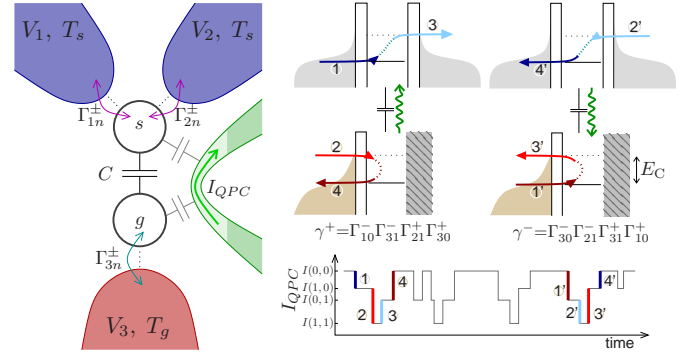}
\caption{\label{sys} Proposed scheme for heat detection. Two capacitively coupled quantum dots are connected to three terminals. One quantum dot ($s$) supports charge transport between two terminals. The other one ($g$) leads to heat transfer from a hot reservoir. A quantum point contact is asymmetrically coupled to both dots serving as a four state charge detector. Time resolved cycles involving all the states, $1{\rightarrow}2{\rightarrow}3{\rightarrow}4$, represent the transfer of one quantum of heat, $E_C$, from the hot to the cold conductor. The reversed sequence, $1'{\rightarrow}2'{\rightarrow}3'{\rightarrow}4'$, takes heat from the cold to the hot reservoir. 
}
\end{figure}

Here we show that charge counting statistics measurements can give access to heat flows and their fluctuations in quantum dot circuits. It allows for tests of non linear fluctuation relations for stationary transport in the presence of thermal and voltage gradients. To illustrate this generally, we consider a model of capacitively coupled conductors for which heat and charge flows are separated: charge flows through the terminals of each conductor; energy is exchanged via electron-electron interactions, as sketched in Fig.~\ref{sys}. Two quantum dots are close enough such that their charge is mutually affected by Coulomb interaction with a capacitance coupling, $C$. Experiments have demonstrated a strong interdot coupling with no electron transfer through the capacitor~\cite{bridge,hubel}. Each quantum dot is coupled by tunneling barriers to different electronic reservoirs considered at local equilibrium. One of them, that we denote as $s$ (for {\it system}), allows charge transport between two terminals at the same temperature but different voltages. The other one, $g$ (for {\it gate}), which is coupled only to one reservoir at a different temperature, supports no stationary charge but a heat current. In previous work, we have shown that such a geometry generates a charge current in the unbiased two terminal conductor by converting the heat flowing though the gate quantum dot~\cite{hotspots,bjorn}. Energy dependent tunneling barriers provide the necessary symmetry breaking~\cite{drag}. The perpendicular directions of heat and charge currents is an important issue for energy harvesting from an environment. Related effects have been predicted in a variety of configurations, when the third hot terminal consists of a phonon bath~\cite{entin,jiang}, the electromagnetic environment~\cite{ruokola2}, incoherent radiation~\cite{rutten} or a spin insulator~\cite{bjorn2}. Recently, current generation from absorption of external radiation has been detected in periodic semiconductor heterostructures~\cite{olbrich,olbrich2}, following ideas by one of us~\cite{markus,blanter} and van Kampen~\cite{vankampen}.

By placing a QPC close to the quantum dots one can determine the charge occupation of the system. The transparency of the constriction is affected by the electrostatic coupling to the charge of the quantum dots. In general, the coupling of the QPC to each quantum dot will be different, strongly depending on their distance. This makes the current flowing through the QPC sensitive to the different charging states. We emphasize that similar arrangements have allowed directional charge counting~\cite{fujisawa,kung} or the detection of quantum operations~\cite{petta} in double quantum dot systems. Alternatively, more complex setups of two QPCs, each one coupled to one quantum dot, can be used~\cite{choi}. The backaction of the detector on the system charge states can be tuned to be negligible. 

We propose detection of energy flows based on a {\it state resolved} charge counting as described in the previous paragraph. We consider a QPC detector with no backaction on the system. In our case, we will focus on a Coulomb blockade regime where each quantum dot is described by a single discrete level that can be occupied by up to one electron. That reduces our  configuration to a four state basis: $(n_s,n_g){=}\{(0{,}0),(1{,}0),(0{,}1),(1{,}1)\}$.  The generalization to cases with several levels is straightforward. Up to sequential tunneling, three states only are not enough to induce correlations between the two conductors that give rise to a noise induced current~\cite{drag}. Due to level discretization, interdot energy flows occur in quanta, $E_C{=}2q^2/\tilde C$, defined in terms of the geometrical capacitance of the coupled system, $\tilde C$~\cite{hotspots}.  Sequences can be identified in the time trace of the current through the QPC, $(0{,}0){\rightarrow}(1{,}0){\rightarrow}(1{,}1){\rightarrow}(0{,}1){\rightarrow}(0{,}0)$, that represent the transfer of a {\it quantum} of energy $E_C$ from the hot gate to the conductor, as sketched in Fig.~\ref{sys}. The reversed trajectory transfers energy from the cold to the hot system. Due to the quantization of transferred energy one can connect its fluctuations to those of the charge occupation of the system.


\section{Model}
\label{sec:model}

We are interested in a system where the heat absorption affects the statistics of a charge current which is measured. For this reason, we investigate a system supporting non parallel electronic charge and heat currents. The three terminal device described above is the minimal model of this characteristics. Heat is transferred between the two dots by electron-electron interaction only. A two terminal configuration with heat transfer only is also discussed.
 
 Coulomb interaction is modeled by the capacitances associated to the tunneling barriers, $C_l$ and the interdot coupling, $C$. In a self-consistent treatment, they define the charging energies, $U_{\alpha,n}(\{V_l\})$, of quantum dot $\alpha$ when the occupation of the other dot is $n{=}0,1$~\cite{hotspots}. The quantum of transferred energy is given by the interdot charging energy, $E_C{=}U_{\alpha,1}{-}U_{\alpha,0}{=}2q^2/\tilde C$, with the effective capacitance $\tilde C{=}(C_{\Sigma s}C_{\Sigma g}{-}C^2)/C$, $C_{\Sigma s}{=}C_1{+}C_2{+}C$, and $C_{\Sigma g}{=}C_g{+}C$. Every terminal is assumed to be at thermal equilibrium with voltage $V_l$ and temperature $T_l$. At sufficiently low temperatures, phononic heat transport can be neglected. We will focus on the case that the two terminals of the conductor are at the same temperature, $T_1{=}T_2{=}T_s$, but can be at different voltages. 

\subsection{Master equation}
\label{sec:meq}
The dynamics of our system can be described by a master equation for the probabilities of the combined system states, $\boldsymbol{\rho}$. We consider the sequential regime, where only diagonal elements of the density matrix are important~\cite{cb,averin}. 
Processes in which an electron tunnels out ($+$) or into ($-$) a quantum dot through
junction $l$ with $n=0,1$ electrons in the other quantum dot are described by the tunneling rates $\Gamma_{ln}^\pm$. They are calculated by Fermi's golden rule, giving: $\Gamma_{\!ln}^-{=}\Gamma_{\!ln}f[(E_{\alpha n}{-}qV_l)\beta_l]$,
$\Gamma_{\!ln}^+{=}\Gamma_{\!ln}{-}\Gamma_{\!ln}^-$, with $\beta_l{=}(kT_l)^{-1}$, $E_{\alpha n}{=}\varepsilon_\alpha{+}U_{\alpha n}$, $f(x){=}(1{+}e^x)^{-1}$ being the Fermi function. $\varepsilon_\alpha$ is the bare energy of the discrete level in quantum dot $\alpha$.
Tunneling rates are in general energy dependent~\cite{saitoh,maclean,astley}. In particular, they depend on the charge occupation of the coupled dot, $n=n_s,n_g$. In the regime $kT_l{\gg}\Gamma_{ln}$, broadening of the energy levels can be neglected.
Written in matrix form, such that $\boldsymbol{\rho}{=}(\rho_{00},\rho_{10},\rho_{01},\rho_{11})^T$, the master equation reads $\boldsymbol{\dot\rho}{=}\boldsymbol{\cal M}\boldsymbol{\rho}$, with
\begin{eqnarray}\label{MLR}
\displaystyle
{
\boldsymbol{\cal M}\!=\!\!
\left(\begin{array}{cccc}
-\Gamma_{\!s0}^-{-}\Gamma_{\!g0}^- &\Gamma_{\!s0}^+& \Gamma_{\!g0}^+ & 0\\
\Gamma_{\!s0}^- & \ \ -\Gamma_{\!s0}^+{-}\Gamma_{\!g1}^- & 0 & \Gamma_{\!g1}^+\\
\Gamma_{\!g0}^- & 0 & -\Gamma_{\!s1}^-{-}\Gamma_{\!g0}^+ & \Gamma_{\!s1}^+\\
0 & \Gamma_{\!g1}^- & \Gamma_{\!s1}^- & -\Gamma_{\!s1}^+{-}\Gamma_{\!g1}^+\\
\end{array}  \right)
}
\end{eqnarray}
and $\Gamma_{\!sn}^\pm{=}\Gamma_{\!1n}^\pm{+}\Gamma_{\!2n}^\pm$. 
It is convenient to separate $\boldsymbol{\cal M}$ into its diagonal and non diagonal terms: $\boldsymbol{\cal M}{=}\boldsymbol{{\cal M}_D}{+}\sum_{l,n,k}\boldsymbol{{\cal J}_{ln}^k}$. The matrices $\boldsymbol{{\cal J}_{ln}^k}$ correspond to those jumps described by tunneling rates $\Gamma_{ln}^k$. They allow one to express the charge, energy and heat stationary currents, $I_{\rm{C},l}$, $I_{\rm{E},l}$ and $I_{\rm{H},l}$, respectively, in a compact form:
\be
\label{currs}
\langle I_{a,l}\rangle{=}\sum_n\text{tr}[{\Theta_{ln}^a}{\left(\boldsymbol{{\cal J}_{ln}^+-{\cal J}_{ln}^-}\right)}\boldsymbol{\bar\rho}],
\ee
where $\Theta_{l,n}^a$ represents the transported amount in each tunneling event: $\Theta_{ln}^{\rm C}{=}q$ (for charge), $\Theta_{ln}^{\rm E}{=}E_{\alpha n}$ (for energy) and
$\Theta_{ln}^{\rm H}{=}E_{\alpha n}{-}qV_l$ (for heat). Note that they are related by: $I_{\rm{H},l}{=}I_{\rm{E},l}{-}I_{\rm{C},l}V_l$. We obtain $\boldsymbol{\bar\rho}$ from the steady state solution given by 
$\boldsymbol{{\cal M}\bar\rho}=0$. From charge conservation, we know that $I_{\rm{C},g}{=}0$ and $I_{\rm{C},2}{=}{-}I_{\rm{C},1}$. Thus  we define $I_{\rm{C}}=I_{\rm{C},2}$ as the total charge current flowing through the conductor. The state resolved charge currents, $\langle I_{ln}\rangle$, are given by the terms inside
the sum in Eq.~\eqref{currs}. 

With this notation, local detailed balance of each reservoir can be expressed as:
\be
\label{detbal}
\boldsymbol{{\cal J}_{ln}^+}=e^{\Theta_{ln}^{\rm H}\beta_l}(\boldsymbol{{\cal J}_{ln}^-})^T, 
\ee
in terms of heat quantities.

\section{Counting statistics and fluctuation theorems}
Briefly, the counting statistics of a vector variable $\boldsymbol{Q}$ is determined by the probability distribution, $P(\boldsymbol{Q},t)$. It is equivalent to consider its cumulant generating function, ${\cal F}(i\boldsymbol{\xi}){=}\ln\sum_{\boldsymbol{Q}}P(\boldsymbol{Q},t)e^{-i\boldsymbol{Q}\boldsymbol{\xi}}$, with counting fields $\boldsymbol{\xi}$. Applied to our system, one can introduce different counting fields $\boldsymbol{\xi}_a{=}\{\xi_{a,l}\}$ (with $a{=}\rm{C,E,H}$) for charge, energy and heat transferred through terminal $l$. Our master equation is modified by introducing the counting fields to multiply the jump operators~\cite{bagrets}:
\be
\boldsymbol{{\cal M}}_{\boldsymbol{\xi},a}{=}\boldsymbol{{\cal M}_D}{+}\sum_{l,n,k}e^{ik\Theta_{l,n}^a\xi_{a,l}}\boldsymbol{{\cal J}_{ln}^k}.
\ee
The cumulant generating function is obtained from the eigenvalue closest to zero of $\boldsymbol{{\cal M}_\xi}$. 
Using \eqref{detbal}, one can easily verify the symmetries of the characteristic polynomials ${\det}{\mid}\boldsymbol{{\cal M}}_{\boldsymbol{\xi},a}{-}\lambda\boldsymbol{\mathbb{1}}{\mid}{=}0$, which translate to different fluctuation theorems. For {\it charge} transport and uniform temperatures, $T_l{=}T, \forall l$, one gets:
\be
\label{fthch}
{\cal F}(i\boldsymbol{\xi_{\rm C}}){=}{\cal F}(-i\boldsymbol{\xi_{\rm C}}+\boldsymbol{A_{\rm C}}),
\ee
with the affinities $A_{\rm{C},l}{=}V_l\beta_l$~\cite{tobiska}. At zero applied voltage, one expects that detailed balance restores equilibrium. However, a finite charge current flows in the presence of hot spots~\cite{hotspots,bjorn}. This apparent broken detailed balance is restored when introducing the energy counting fields as well~\cite{andrieux2}. 
Alternatively, a more compact form can be written in terms of {\it heat} currents:
\be
\label{heatfth}
{\cal F}(i\boldsymbol{\xi_{\rm H}}){=}{\cal F}(-i\boldsymbol{\xi_{\rm H}}{-}\boldsymbol{A_{\rm H}}),
\ee
with the energy affinities $A_{{\rm H},l}{=}\beta_l$. We recall that heat currents are related to charge and energy currents. It is therefore valid in the presence of both voltage and temperature gradients. 

We show below that heat current statistics are obtained by state resolved electron counting if the energies $\Theta_{l,n}^{\rm E}$ are known. 
If we are interested in the heat transferred between the two conductors, only the difference $\Theta_{l,1}^{\rm E}{-}\Theta_{l,0}^{\rm E}{=}E_C$ becomes relevant, which can be accessed experimentally in our configuration. A complete level resolved detection including every terminal requires a more complicated setup.

\subsection{State resolved fluctuation relations}
For state resolved counting one introduces the number of transferred particles through terminal $l$ and from state $n$, $N_{ln}$, and the vector of level resolved counting fields, $\boldsymbol{\zeta}{=}\{\zeta_{ln}\}$. Then the jump operators in the master equation are multiplied by $\exp[ik\zeta_{ln}]$. The explicit level dependence of local detailed balance conditions in Eq.~\eqref{detbal} leads to the state resolved fluctuation relation:
\be
\label{frellev}
{\cal F}(i\boldsymbol{\zeta}){=}{\cal F}(-i\boldsymbol{\zeta}{-}\boldsymbol{\tilde A}),
\ee
where $\tilde A_{l,n}{=}\Theta_{l,n}^{\rm H}\beta_l$ are the state resolved affinities. 
In contrast to Eq.~\eqref{fthch}, the relation shown in Eq.~\eqref{frellev} is a fluctuation relation for particles which is valid in the presence of voltage and temperature gradients, as is the case for the heat fluctuation theorem~\eqref{heatfth}. 
As a drawback, it is no longer universal but depends on the internal configuration of the system --- in particular, on gate voltages. However, we stress that, in a self-consistent treatment, $\Theta_{l,n}^{\rm H}{=}E_{\alpha,n}(\{V_m\}){-}qV_l$ depends only on differences of voltages, so Eq. \eqref{frellev} is gauge invariant.

\subsection{Two terminals: Pure heat transport}
\label{sec:diode}

\begin{figure}
\center
\includegraphics[width=52mm,clip]{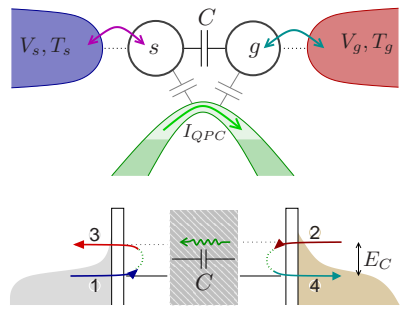}
\caption{\label{diode} Heat diode in a two terminal configuration. Here, no charge but only heat flows through the capacitor.  
}
\end{figure}

Let us first investigate a simple configuration that illustrates the above result. We consider a two terminal system where no stationary charge current will flow~\cite{ruokola}. A schematic representation of such a system with the QPC detector added is shown in Fig.~\ref{diode}. Note that in this configuration, heat and energy currents coincide, $I_{\rm{H},l}{=}I_{\rm{E},l}{=}E_CI_{l1}/q$, and therefore are conserved: $\sum_lI_{\rm{H},l}{=}\sum_lI_{\rm{E},l}{=}0$.

This configuration allows for complete state resolved counting. The quantity of interest is the heat flowing through the gate. This is independent of the number of levels of each dot and is determined solely by the charging energy. The relevant index is therefore given by the charge occupation. For quantum dots that permit up to one electron, $n{=}1,0$ labels the charge occupation of the dot that is not involved in the tunneling event. The gate $g$ releases a quantum of energy $E_C$ to the system $s$ when a cycle (depicted in Fig.\ref{diode}) is completed: an electron enters dot $g$ when dot $s$ is occupied and leaves it only once the dot $s$ is empty. Thus, the amount of energy transferred to the gate lead is proportional to the difference of the number of electrons that have tunneled from the two states $(1{,}1)$ and $(0{,}1)$: 
\be
\label{eg}
\tilde E_g{=}E_C(N_{g1}{-}N_{g0})/2.
\ee
This equation relates the transferred heat statistics to the charge fluctuations in the gate, which can be measured in a state dependent detector as the one shown in Fig.~\ref{diode}.

In this case, the validity of the fluctuation theorem can be verified by detecting the statistics of charge transferred from a given energy. The constraints introduced by charge conservation in each conductor ($N_{l0}{=}{-}N_{l1}$) and energy conservation ($N_{s1}{=}{-}N_{g0}$) permits to write separate relations for particles emitted through any state or terminal, $N_{ln}$. 
For example, one can simply count the number of electrons $N_{sn}$ emitted to the cold system conditioned on the hot dot containing $n$ electrons. The corresponding cumulant generating function obeys the universal symmetry:
\be
\label{diodeFth}
{\cal F}(i\zeta_{sn}){=}{\cal F}(-i\zeta_{sn}{+}E_C\left(\beta_g{-}\beta_s\right)).
\ee
For particles emitted through the gate terminal, replace $s\leftrightarrow g$ in the previous expression.
No gate voltage dependence enters Eq.~\eqref{diodeFth}. 

\subsection{Three terminals: Noise induced transport}
We are mostly interested in the deviations of charge current statistics in the presence of heat absorption. For that we consider a three terminal device, as the one described above and shown in Fig.~\ref{sys}. 
In a recent work, Krause {\it et al.}~\cite{krause} showed that state resolved counting of electrons allows fluctuation relations to be written when the statistics of dissipated energy is also measurable. It is however challenging to do so for phonons.
In our setup, on the contrary, the energy absorbed by a conductor is only due to electron-electron interactions and can be treated on the same footing as the particle currents, cf. Eq.~\eqref{eg}. 

Charge and energy conservation can be written in terms of state resolved counting: ${\sum_{l\in s,n}}N_{ln}{=}{\sum_n}{N}_{g}{=}0$ and ${\sum_{l\in s}}(N_{l,1}{-}N_{l,0}){=}N_{g,1}{-}N_{g,0}$, respectively. They provide additional symmetries of the cumulant generating function expressed as shifts of the counting fields. Separating the contributions from each conductor, system and gate, $\boldsymbol{\zeta}{=}(\boldsymbol{\zeta_s},\boldsymbol{\zeta_{g}})$, we write:
\bea
\label{enercons}
&&{\cal F}(i\boldsymbol{\zeta}_{\boldsymbol{s}0},i\boldsymbol{\zeta}_{\boldsymbol{s}1},i\boldsymbol{\zeta}_{\boldsymbol{g}0},i\boldsymbol{\zeta}_{\boldsymbol{g}1})\nonumber\\
&{=}&{\cal F}(i\boldsymbol{\zeta}_{\boldsymbol{s}0}{+}a,i\boldsymbol{\zeta}_{\boldsymbol{s}1}{+}a,i\boldsymbol{\zeta}_{\boldsymbol{g}0}{+}b,i\boldsymbol{\zeta}_{\boldsymbol{g}1}{+}b)\\
&{=}&{\cal F}(i\boldsymbol{\zeta}_{\boldsymbol{s}0}{-}c,i\boldsymbol{\zeta}_{\boldsymbol{s}1}{-}d,i\boldsymbol{\zeta}_{\boldsymbol{g}0}{+}d,i\boldsymbol{\zeta}_{\boldsymbol{g}1}{+}c),\nonumber
\eea
for arbitrary parameters $a$, $b$, $c$ and $d$. The dimension of vectors $\boldsymbol{\zeta_{s}}$ and $\boldsymbol{\zeta_{g}}$ is given by the number of levels in each conductor times the number of terminals to which it is connected.
If the states of the gate are not resolved, only electrons flowing through the conductor are measured,
Eqs.~\eqref{enercons} lead to an incomplete fluctuation theorem~\cite{krause}:
\be
\label{fs}
{\cal F}(\{i\zeta_{sn}\}){=}{\cal F}(\{-i\zeta_{sn}{-}\tilde A_{sn}{-}\tilde A_{g\bar n}\}).
\ee
Note that the system and gate affinities are taken at different occupations, $\bar n{=}\text{not}(n)$. It applies to cases where the QPC is not able to distinguish the states $(0{,}0)$ and $(0{,}1)$: for example if the difference of the corresponding signals is smaller than the noise of the  measurement.

The state resolved cumulant generating function is defined in terms of the probability to detect $\mathbf N$ transferred electrons during a period of time, $t$: ${\cal F}(i\boldsymbol{\zeta}){=}\ln\sum_{\boldsymbol{N}}P(\boldsymbol{N},t)e^{-i\boldsymbol{N}\boldsymbol{\zeta}}$. For later convenience, the charge currents will be written in terms of the number of electrons transferred in a given time: $I_{ln}{=}qN_{ln}/t$.
Then, in the long time limit, Eq.~\eqref{fs} can be expressed in terms of the probability to measure a given current configuration, $\boldsymbol{I_{s}}=\{I_{ln}\}$, $l{\in}s$, in the conductor~\cite{espositorev}:
\be
\frac{q}{t}{\ln}\frac{P(\boldsymbol{I_{s}})}{P(-\boldsymbol{I_{s}})}{=}{\sum_{l\in s,n}}I_{ln}(\tilde{A}_{l,n}{+}\tilde{A}_{g,\bar n}).
\ee
In the case of interest here where the two terminals of the conductor $s$ are at the same temperature, the previous expression can be notably rewritten in terms of 
the total charge current, $I_{\rm{C}}{=}q\sum_nN_{2n}/t$, and the amount of energy absorbed from the gate, $I_{\rm{H},g}{=}\tilde E_g/t$, expressed in terms of particle flows in the conductor, cf. Eq.~\eqref{eg}:
\be
\label{pn}
\frac{1}{t}{\ln}\frac{P(\boldsymbol{I_{s}})}{P(-\boldsymbol{I_{s}})}{=}I_{\rm{C}}(V_1{-}V_2)\beta_s{+}I_{\rm{H},g}(\beta_g{-}\beta_{s}).
\ee
It resembles fluctuation theorems for thermoelectric transport that separate thermal and electric forces~\cite{andrieux2,campisi,krause,simine}. However, Eq.~\eqref{pn} is configuration dependent: the right hand side depends on state resolved currents through $I_{\rm{H},g}$. For instance, it is affected by gate voltages. For single level quantum dots, it reads: $I_{\rm{H},g}{=}-E_C(I_{\rm{C}}{+}I_{11}{-}I_{20})/q$, where the last two terms represent the difference of state resolved currents $I_{ln}$ flowing through the different leads ($l{=}1,2$) of the conductor at the different charge occupations ($n{=}1,0$) of the gate quantum dot. 

\begin{figure}
\center
\includegraphics[width=76mm,clip]{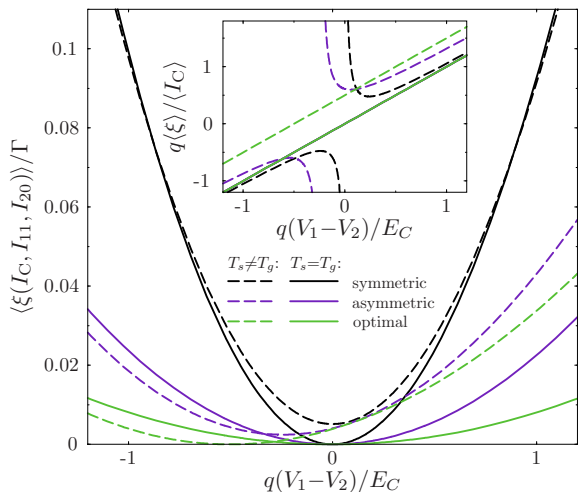}
\caption{\label{exp} Fluctuation relation. We plot the right hand side of Eq.~\eqref{pn} written in terms of mean charge currents: $\langle\xi\rangle{=}\langle I_{\rm{C}}\rangle(V_1{-}V_2)\beta_s{-}E_C(\langle I_{{\rm C}}\rangle{+}\langle I_{11}\rangle{-}\langle I_{20}\rangle)(\beta_g{-}\beta_s)/q$, as a function of the bias voltage applied to the conductor. Different tunneling rate configurations are shown for homogeneous ($T_s{=}T_g$) and inhomogeneous ($T_s{\ne}T_g$) temperatures: ``symmetric"($\Gamma_{ln}{=}\Gamma,\ \forall\{l,n\}$), ``asymmetric"($\Gamma_{ln}{=}\Gamma$, except $\Gamma_{11}{=}\Gamma_{20}{=}\Gamma/10$), and ``optimal"($\Gamma_{ln}{=}\Gamma$, except $\Gamma_{11}{=}\Gamma_{20}{=}0$). The temperature of the conductor is kept $kT_s{=}5\hbar\Gamma$, while the gate is heated to $kT_g{=}10\hbar\Gamma$ in the inhomogeneous case. The presence of a hot spot generates a shift of the minimum. It does not cross the origin except for the optimal case, in which it goes to zero at the stall potential of Eq.~\eqref{stall}. The inset shows the same quantity $\langle\xi\rangle$ but normalized to the total charge current, $\langle I_{\rm{C}}\rangle$, so all the isothermal configurations have the same slope. The absence of detailed balance gives a divergence at the stall voltage.
}
\end{figure}

An illustration of the previous expression is plotted in Fig.~\ref{exp}. We denote the right hand side of Eq.~\eqref{pn} with $\xi$. Remember that $I_{\rm{H},g}$ can be written in terms of charge currents. 
In equilibrium, $\xi{=}0$.
Remarkably, in asymmetric configurations that manifest noise induced transport, $\langle\xi\rangle$ does not vanish even if the total current, $\langle I_{\rm{C}}\rangle$, does at a certain stall potential. Though at that voltage no net charge currents flow, heat flowing through the gate drives the system out of equilibrium. Detailed balance is restored only at the optimal converter configuration, cf. Eq.~\eqref{pnun} below.

Optimal heat to charge conversion occurs in particular cases where the tunneling events $(0{,}0){\leftrightarrow}(1{,}0)$ and $(0{,}1){\leftrightarrow}(1{,}1)$ can only take place through different terminals of the conductor~\cite{hotspots}. For instance, if $\Gamma_{11}{\ll}\Gamma_{10}$ and $\Gamma_{20}{\ll}\Gamma_{21}$, so $I_{11},I_{20}{\rightarrow}0$. Then, a universal relation holds:
\be
\label{pnun}
\frac{1}{t}{\ln}\frac{P(I_{\rm{C}})}{P(-I_{\rm{C}})}{=}I_{\rm{C}}{\left[{(V_1{-}V_2)}\beta_s{-}\frac{E_C}{q}{(\beta_g{-}\beta_{s})}\right]}.
\ee
The cycles involved are illustrated in Fig.~\ref{sys}. At this configuration, a charge $q$ is transported across the conductor for each quantum of heat $E_C$ that is exchanged between the two systems, so $I_{\rm{C}}/q{=}I_{\rm{H},g}/E_C$. This implies that $I_{\rm{C}}$ and $I_{\rm{H},g}$ are maximally correlated, i.e. $\langle I_{\rm{C}}I_{\rm{H},g}\rangle{=}\langle I_{\rm{C}}\rangle\langle I_{\rm{H},g}\rangle$, and all their cumulants coincide. 

Note that a related expression can be obtained using the relation of rates for the cycle that transfers an energy $E_C$ to the conductor, $\gamma^+{=}\Gamma_{10}^-\Gamma_{31}^-\Gamma_{21}^+\Gamma_{30}^+$, and its reversed sequence, $\gamma^-{=}\Gamma_{30}^-\Gamma_{21}^-\Gamma_{31}^+\Gamma_{10}^+$, which gives:
\be
\label{g+-}
\frac{\gamma^+}{\gamma^-}{=}\exp[q(V_1{-}V_2)\beta_s{+}E_C(\beta_s{-}\beta_g)].
\ee
The exponent on the right hand side of Eq.~\eqref{g+-} is given by the entropy flux per cycle~\cite{seifert,esposito}. We can therefore interpret the right hand side of Eq.~\eqref{pnun}: it relates the entropy flux to the Joule heat in the conductor and heat absorbed from the gate. It also provides the condition for the stall potential
\be
\label{stall}
q(V_1{-}V_2){=}E_C\frac{\beta_g{-}\beta_s}{\beta_s}{=}{-}E_C\eta_C,
\ee
where a conversion efficiency approaching the Carnot limit is predicted~\cite{hotspots}.
This stall potential defines a non linear analogue of a Seebeck thermopower, $E_C/(qT_g)$, for noise induced transport.

\section{Conclusions}
\label{sec:conclusions}

To summarize, we propose a mechanism to detect the statistics of electronic heat currents in Coulomb blockaded quantum dot systems. Multilevel charge detection by means of a side coupled quantum point contact allows to count the transfer of discrete energy $E_C$.  The same scheme permits to investigate deviations of charge flow statistics in the presence of hot spots, manifested in non universal fluctuation relations. We investigate  a three terminal heat to charge converter where noise induced transport avoids detailed balance of charge flows. Only for an optimal configuration when the absorption of a quantum of heat $E_C$ leads to the transfer of a charge $q$ across the conductor, can a fluctuation theorem for total charge flow be written. In a two terminal device with no net charge currents, state resolved charge counting statistics fulfill a universal fluctuation theorem equivalent to the one obtained for heat currents. All our results can be probed within present experimental technology.

\acknowledgements
We thank A. N. Jordan and B. Sothmann for a critical reading of the manuscript. Work supported by the Spanish MAT2011-24331, the EU ITN Grant 234970, the Swiss NSF and the EU project NANOPOWER (FP7/2007-2013) under grant agreement no. 256959.  R. S. acknowledges support by the CSIC and FSE JAE-Doc program.

\end{document}